\documentclass[aps,prb,showpacs,twocolumn,superscriptaddress]{revtex4-1}

\usepackage{graphicx}
\usepackage{amsmath,amssymb}
\usepackage{hyperref}
\usepackage{natbib}
\usepackage{bm}

\begin{document}


\title{Periodically driven interacting electrons in 1D: a many-body  Floquet approach}

\date{\today}
\author{M.~Puviani}
\affiliation{Dipartimento di Scienze Fisiche, Informatiche e Matematiche, Universit\`a di Modena e Reggio Emilia, Via Campi 213/A, I-41125 Modena, Italy }\author{F.~Manghi}
\affiliation{Dipartimento di Scienze Fisiche, Informatiche e Matematiche, Universit\`a di Modena e Reggio Emilia, Via Campi 213/A, I-41125 Modena, Italy }
\affiliation{CNR - Institute of NanoSciences - S3}

\begin{abstract}

We propose a method to study the time evolution of correlated electrons driven by an harmonic perturbation.  Combining Floquet formalism to include the time-dependent field and  Cluster Perturbation Theory to solve the many-body problem in the presence of short-range correlations, we treat the electron double dressing - by photons and by e-e interaction -  on the same footing. We apply the method  to an extended Hubbard chain at half occupation and we show that in the regime of small field frequency  and for given values of field strength the zero-mode Floquet band is no more gapped and the system recovers a metallic state.  Our results are indicative of an omnipresent  mechanism for insulator-to-metal transition in 1D systems.

\end{abstract}

\maketitle

Under the influence of periodic fields quantum systems may reach regimes inaccessible under equilibrium conditions and new phases may be engineered by an external tunable control.
The recent experimental evidence of optically quenched superconductivity  \cite{Zhang2014}   or  the possibility to induce topological phases by light irradiation in systems that would be standard in stationary conditions \cite{Sentef2015,Ezawa2013,Cayssol2013b,Lindner2011,Oka2009} are  relevant examples in this field.

The coexistence of periodic external driving forces and  electron-electron correlations is particularly interesting and this for two main reasons: on one side, according to the Peierls' substitution \cite{Hofstadter1976}, the external field effectively reduces  the inter-site hopping so  enhancing the  effects of the e-e repulsion and the tendency to an insulating behaviour.  On the other side irradiation may be responsible of a photo-doping that may affect  the charge carrier density and turn a Mott insulator  into a metal \cite{Okamoto2007,Iwai2003}. Due to these two competing effects novel
phenomena are expected when  strongly correlated  quantum systems are exposed to  time-dependent external fields.

In this letter we explore the combined effects of a time-periodic field and of  on-site e-e interaction on an extended 1D lattice. The 1D Hubbard chain in static conditions is  the prototype of Mott-Hubbard insulators, exhibiting at half filling an insulating behaviour no matter how weak the e-e repulsion is. In the presence of a continuous periodic driving the system is in a non-equilibrium steady state characterized  by a periodic time dependence. In this situation
the Floquet formalism \cite{Shirley1965,Aoki2014} can be adopted that converts  the time-dependent problem into a time-independent one: the application of the external  field transforms  electronic states  into  photon-dressed   quasi-energies where virtual absorption and emission of photons is taken into account to all orders.
In most cases the Floquet formalism is applied to non-interacting systems \cite{Faisal1997,Cayssol2013b,Usaj2014} and in order to extend it to the interacting case it must be reformulated in terms of the one-particle Floquet Green's function (GF) whose imaginary part will then describe a double dressing,  by e-e interaction and by photons.

We have developed a method to calculate the Floquet GF for extended lattices based on the Cluster Perturbation Theory (CPT) \cite{Senechal}. CPT  belongs to the class of Quantum Cluster theories\cite{RevModPhysQC} that solve the problem of many interacting electrons in an extended lattice by a  divide-and-conquer  strategy, namely by solving first the many body problem in a subsystem of finite size and then embedding it within  the infinite medium.
CPT has been successfully used to study Mott-Hubbard physics both in model systems \cite{Senechal2000,Potthoff2003} and in real materials \cite{Manghi2014,Eder2008,Eder2015}, and more recently to address correlated topological phases of matter. \cite{Grandi.NJP,Grandi.PRB,PhysRevLett.107.010401,Wu2006}

The solution of the many-body Hamiltonian in a finite cluster is the starting point.
The Floquet time-independent eigenvalue problem extended to a many-body Hamiltonian (the Hubbard model in our case) reads
\begin{equation}
\label{floqueteigen}
\hat{H}^F \Phi^N_{\alpha}(X,t) = E^N_{\alpha} \Phi^N_{\alpha}(X,t) \ ; \  X\equiv (x_1,x_2,...,x_N).
\end{equation}
It involves the Floquet-Hubbard Hamiltonian
\begin{equation}\label{FloqH}
    \hat{H}^F  \equiv \left[ \hat{H}  -\imath \frac{\partial}{\partial t}\right]
\end{equation}
 with
\begin{equation}\label{H}
\hat{H}=  \sum_{ii'\sigma} \tilde{J}_{i i'}(t)\hat{c}_{i \sigma}^{\dag}(t) \hat{c}_{i'\sigma}(t) +
U \sum_{i} \hat{c}_{i \uparrow}^{\dag}(t)\hat{c}_{i \uparrow}(t)\hat{c}_{i \downarrow}^{\dag}(t)\hat{c}_{i   \downarrow}(t)
\end{equation}
Here $\tilde{J}_{i i'}(t)$ describes the hopping
between neighboring sites modified by the external vector potential $A(t)=A_0 sin(\Omega t)$ according to  Peierls' substitution \cite{PhysRevLett.108.225303,Hofstadter1976}
\begin{equation}\label{hop}
   \tilde{J}_{i i'}(t)=J e^{iA(t)\cdot (r_{i'}-r_i)}
\end{equation}
$J$ being the unperturbed nearest neighbour hopping.

The solution of Eq. (\ref{floqueteigen}) provides the time-periodic part  of the many-body wavefunction
\begin{equation}
\label{eq2} \Psi^N_{\alpha}(X,t)= e^{-\imath E^N_{\alpha}t} \Phi^N_{\alpha}(X,t) \ ; \ \Phi^N_{\alpha}(X,t)=\Phi^N_{\alpha}(X,t+nT)
\end{equation}
Being $\Phi^N_{\alpha}(X,t)$ time-periodic it can be expressed as a Fourier series
\begin{equation}\label{FT}
  \Phi^N_{\alpha}(X,t)=\sum _{n=-\infty}^{\infty} A^N_{\alpha n}(X) e^{-\imath n \Omega t} \  ; \Omega\equiv \frac{2 \pi}{T}.
\end{equation}
where in turn $A^N_{\alpha n}(X)$ can be expanded on a complete set of non-interacting N-particle wavefunctions.  In practice the summation is truncated to include a finite number of modes, up to a sufficiently large $n_{max}$ whose value depends obviously on $\Omega$.

For a finite system this set of equations is solved  by exact diagonalization   \cite{Eckardt2005,Takahashi2008,Grushin2014,Mentink2015}  and its
eigenstates are used to obtain the Floquet GF   in terms of hole and particle propagators
\onecolumngrid
\begin{eqnarray}\label{FGFFT}
  G{\substack{i i' \\
nn'}} (\omega)&=&
\sum_{\substack{m m' m'' m''' \\
\alpha}} \left(
 \frac{<A^N_{0 m}|\hat{c}^{\dag}_{i n}|A^{N-1}_{\alpha m'}><A^{N-1}_{\alpha m''}|\hat{c}_{i n'}|A^{N}_{0 m'''}>   }{\omega-(E_0^{N}-E^{N-1}_{\alpha})-i\eta} \delta(m'+n-m)\delta(m'''-n'+m'') \right.\\ \nonumber
&+& \left. \frac{<A^N_{0 m}|\hat{c}_{i n}|A^{N+1}_{\alpha m'}><A^{N+1}_{\alpha m''}|\hat{c}^{\dag}_{i n'}|A^{N}_{0 m'''}>   }{\omega-(E^{N+1}_{\alpha}-E^{N}_{0})+i\eta} \delta(m'-n-m)\delta(m'''+n'-m'')\right)
\end{eqnarray}
\twocolumngrid
Here $i,i'$ indicate sites and $n,n'$ the  Floquet modes appearing in eq.  (\ref{FT} ).
Eq. (\ref{FGFFT}) is  the Lehmann representation of the Floquet GF  for a finite system; it can be explicitly calculated in terms of the few-body eigenstates obtained by exact diagonalization for $N$ and $N\pm 1$ electrons.

Notice that the two delta-functions  that appear in eq. (\ref{FGFFT}) express  a  conservation rule on Floquet modes: removing/adding  mode $m$  from/to the $N$-particle state  gives rise to a $N^-/N^+$ state  where each Floquet component has been depleted/augmented by the same quantity.

The definition of the Floquet ground state $E^{N}_{0}$   in eq. (\ref{FGFFT}) requires some further comments. Floquet eigenstates are auxiliary quantities  that allow to calculate the  time-dependent eigenfunctions  $\Psi^N_{\alpha}(X,t)$ and from them the  time-dependent energies of the true Hamiltonian  as
\begin{equation}\label{Et}
   E^N_{\alpha}(t)= <\Psi^N_{\alpha}(X,t)|H(t)|\Psi^N_{\alpha}(X,t)>
\end{equation}
The Floquet ground state may then be identified  as the one that minimizes the time-average of this quantity. However, this definition is acceptable only for nearly-vanishing values of the external field $A_0$ and/or in the high frequency regimes $\Omega \gg J$ where the mixing between Floquet bands is minimum. \cite{Kohn2001,Hone1997} In order to identify the Floquet ground state at any  field intensity we have adopted a \emph{continuity} criterion:  having identified the ground state in the limit of very small field intensity according to the time-averaged expectation value minimization,   we assume it to vary continuously  as a function of $A_0$ (more details in the  supplemental material).

The Floquet GF for the extended lattice $G^{lat}$ is obtained in terms of Floquet GF for decoupled clusters connected by  inter-cluster hopping.  We get
\begin{equation}\label{FCPT}
G^{lat}_{\substack{i i' \\
nn'}}(k,\omega)=G_{\substack{i i' \\nn'}}(\omega)+ \sum_j B_{\substack{i j \\
nm}}(k  \omega) G_{\substack{j i' \\m n'}}(k   \omega).
\end{equation}

Here $G_{\substack{i i' \\nn'}} (\omega)$  is the cluster Floquet GF of eq. (\ref{FGFFT}) and the matrix $B_{\substack{i i' \\ nn'}}(k,\omega)$ is the generalization to Floquet space of the analogous matrix defined in CPT \cite{Manghi2014,Grandi.NJP,Grandi.PRB}
\begin{align}
\label{B}
B_{\substack{i i' \\
nn'}}(k,\omega)= \sum_l e^{i k\cdot R_l} \sum_{i'' n''} G_{\substack{i i'' \\
nn''}}(\omega) \tilde{J}_{n'' n'}.
\end{align}

Finally we get the Floquet spectral function, namely the spectrum of the photon dressed   quasiparticle energies:
\begin{align}
\label{specakn}
D(k,\omega) = \frac{1}{\pi}\sum_m {\rm Im}\,  G^{lat}(k,m,\omega)
\end{align}
where
 \begin{align}
 \label{Glat}
G^{lat}(k,m,\omega) =  \sum_{\substack{i i'  \\
n n'}}  e^{-i k \cdot(r_i-r_{i'})}\alpha^{m*}_{i n}(k) \alpha^m_{i'n'}(k)  G^{lat}_{\substack{i i' \\n n'}}(k,\omega)
 \end{align}
Here $\alpha^{m*}_{i n}(k)$,$ \alpha^m_{i'n'}(k) $ are  the single-particle Floquet eigenvalues obtained by solving the Floquet Hamiltonian for non-interacting electrons. Equations (\ref{FCPT}-\ref{Glat}) represent the  extension  of the standard CPT formalism developed in static conditions \cite{Manghi2014,Grandi.NJP,Grandi.PRB} to the time-dependent case.

The novelty of the present approach consists first of all in the definition of the  Floquet GF for the isolated cluster ( eq. (\ref{FGFFT}) ) and secondly in the CPT embedding procedure that provides  the GF for the extended lattice in terms of cluster GF's. We stress again that two main points are essential in order to get a proper definition of the cluster Floquet GF: the conservation rule  on Floquet modes expressed by the two delta-functions in eq.  (\ref{FGFFT} ) and the proper definition of the Floquet ground state.  The conservation rule on Floquet modes in particular is essential to recover the correct non-interacting limit where the Floquet spectral functions must reproduce exactly the single-particle Floquet band structure.

\begin{center}
\begin{figure}[h]
 \centering \includegraphics[width=18cm]{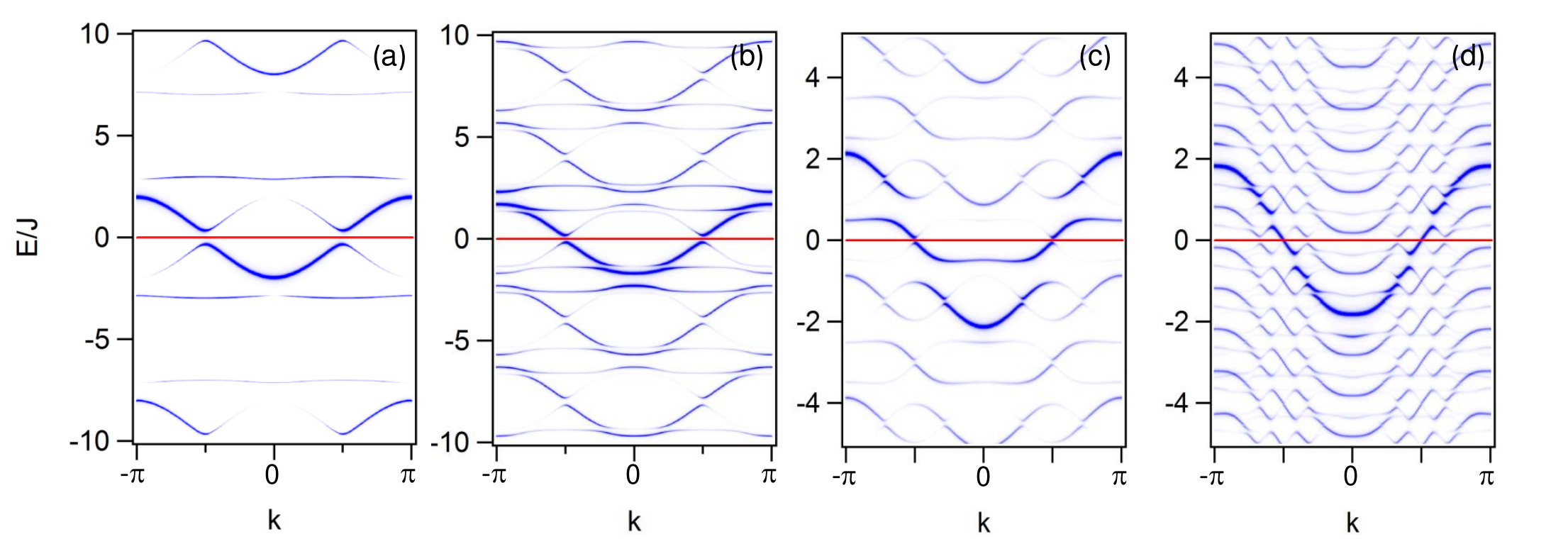}
 \caption{\label{aakknn} (color online) Photon-dressed quasi-particle bands for the extended Hubbard chain with $U/J=2$,  $A_0/J=0.7$ and   different field frequencies: (a) $\Omega=10J$, (b) $\Omega=4J$, (c) $\Omega=3J$, (d) $\Omega=1J$. Darker lines around $E/J=0$ indicate the Floquet zero mode. In the regime of very large frequency the Floquet spectrum consists in equally - spaced replicas of  the k-resolved spectral function obtained in static conditions. }
\end{figure}
\end{center}

Using this new approach we have performed a systematic study of the photon-dressed quasi-particle bands for an extended Hubbard chain at half filling for a given value of $U/J$  in different regimes of field intensity and frequency. Fig. \ref{aakknn} reports the result obtained for small field intensity ($A_0=0.7 J$). Heavier lines identify the Floquet zero mode. We observe that in the limit of very large frequencies (Fig. \ref{aakknn} (a) ) the Floquet spectrum  corresponds, as expected, to replicas of the original quasi-particle band structure spaced by integers multiples of $\Omega$: the Floquet bands are well separated copies of the k-resolved spectral functions obtained in static conditions, characterized by a well defined Mott-Hubbard  gap and by non dispersive satellites structures below and above the valence and conduction band respectively.

\begin{center}
\begin{figure}[h]
 \centering \includegraphics[width=18cm]{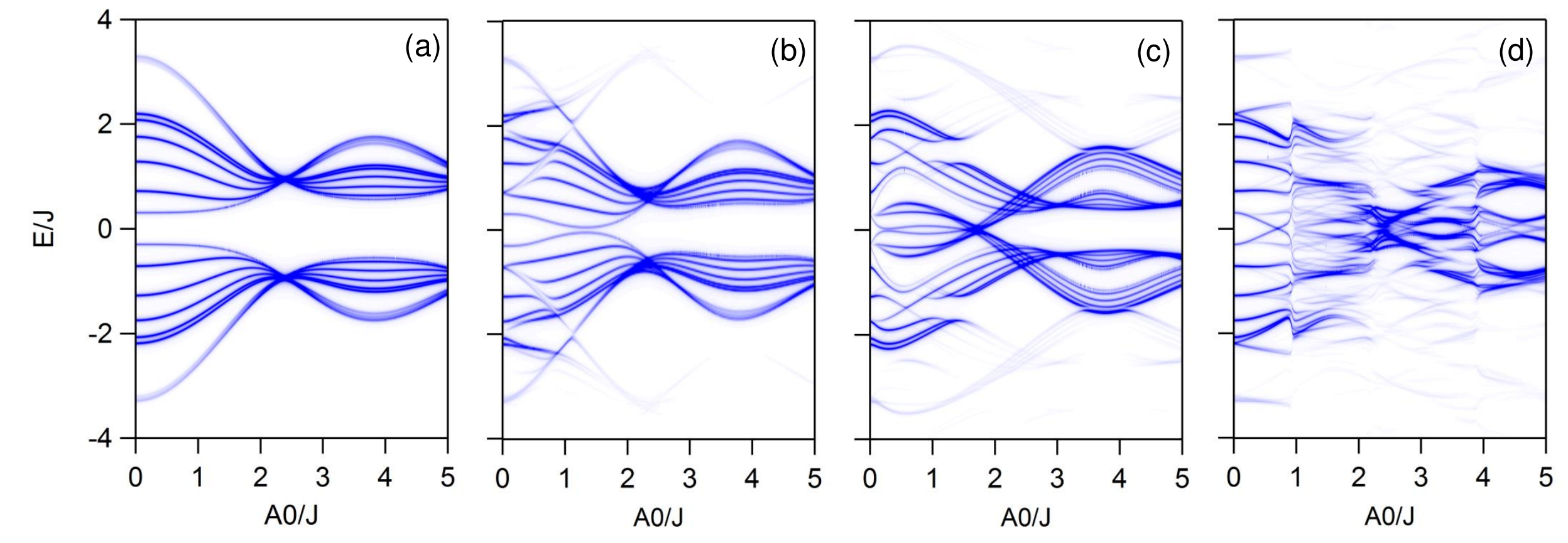}
 \caption{\label{plot} (color online) Zero mode photon dressed quasi-particle bands calculated with $U/J=2$ at 11 k-points in the 1D Brillouine zone reported as a function of $A_0$   in the same frequency regimes of Fig. \ref{aakknn}: (a) $\Omega=10J$,(b) $\Omega=4J$,(c) $\Omega=3J$,(d) $\Omega=1J$. }
\end{figure}
\end{center}

For smaller field frequencies the quasi-particle bands overlap and mix, giving rise to the Floquet spectra  shown  in Fig. \ref{aakknn} (b-d).
A noticeable result is that for  the smallest frequency (Fig. \ref{aakknn} (d) )  the Mott-Hubbard gap is no more present.  This is true not only at this particular field strength but it appears as a constant characteristic of  the low frequency regime. This is shown in  Fig. \ref{plot}  where we report  the evolution of the Floquet zero mode as  a function of filed strength,  from  $A_0=0$ to  $A_0=5J$.
\onecolumngrid
\twocolumngrid
\onecolumngrid
\twocolumngrid
While at large frequencies the Mott gap remains  open at all field strengths, for frequencies smaller than the  static quasi-particle band-width  the gap closes down at some specific values of $A_0$. Interestingly the  regime of very small frequencies (Fig. \ref{plot} (d) ) exhibits the strongest gap modulation and we observe a gap closing not only for  $A_0/J< 1$ but also around $A_0/J= 2.8$ and $A_0/J=3.8 $.
\begin{center}
\begin{figure}[h]
 \centering \includegraphics[width=9cm]{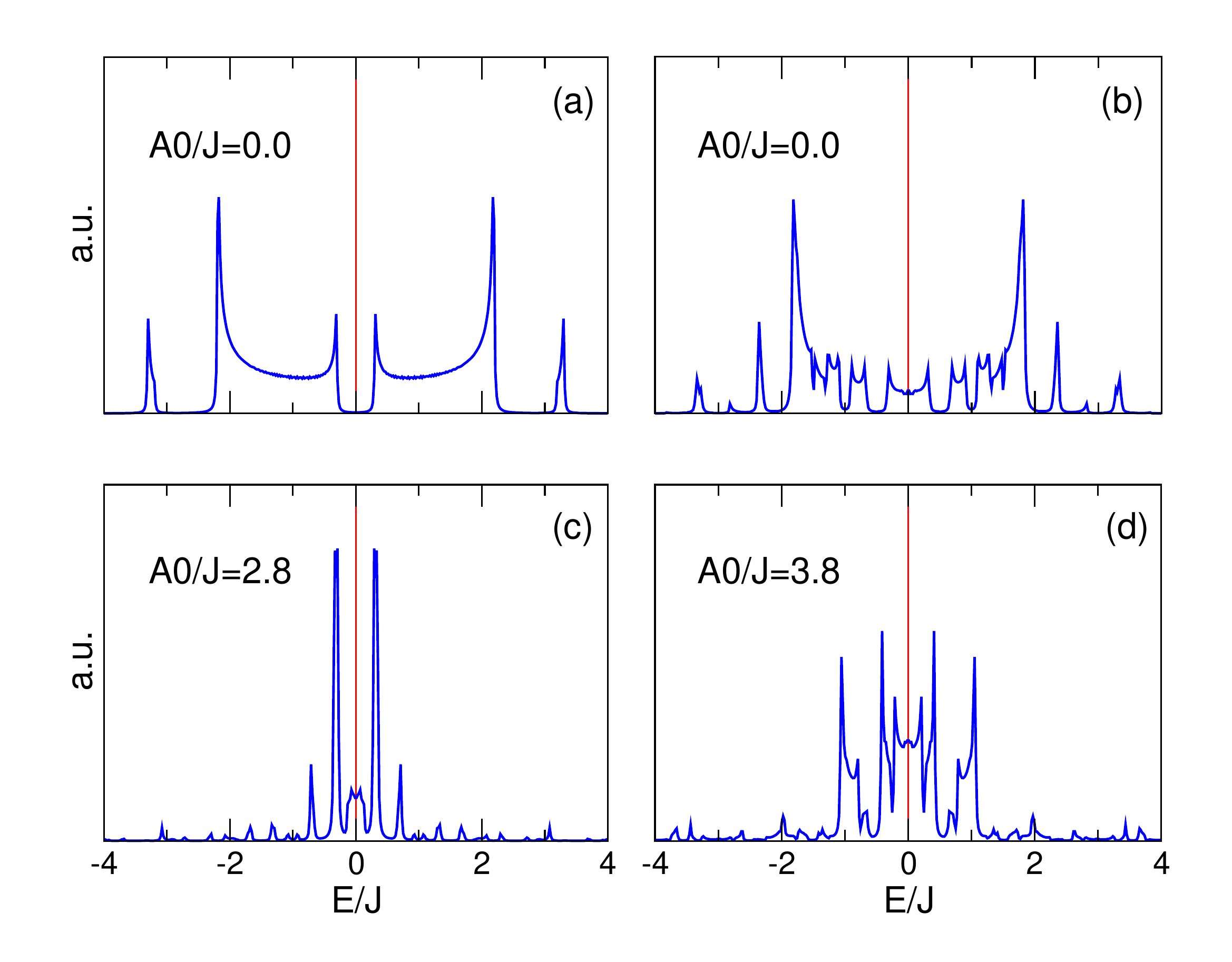}
 \caption{\label{dosc} (color online) Density of zero-mode photon dressed quasi-particle states in arbitrary units for  $U/J=2$, $\Omega/J=1$ for different $A_0$. Panel (a) reports the static result, showing the well defined Mott-Hubbard gap, panels (b)-(c) refer to values of $A_0$ responsible of a metallic phase. }
\end{figure}
\end{center}
It is then clear that  for these field frequencies  an external ac drive of appropriate strength may induce an insulator-to-metal transition. This is clearly shown in Fig. \ref{dosc} where  a gapless  density of photon-dressed quasi-particle states is obtained at these parameter values (Fig. \ref{dosc} ((b)-(d))  as compared with the gapped static result (Fig. \ref{dosc} (a) ).

The ability of time-periodic fields to  turn an insulator into a metal  has been previously shown  for non-interacting electrons in  a  dimerised 1D chain \cite{Gomez-Leon2013}.  The same effect is found here  even if the  physical origin of the insulating phase is  drastically different:   a many-body effect due to local e-e repulsion instead of a single-particle  mechanism due to lattice dimerization.  Since in 1D any system in static conditions is  a Mott  insulator, this result  is ubiquitous and for any 1D material we will be able to photo-induce a metallic state  by choosing an external drive with the correct combination of frequency and intensity. Of course these values depend on the system properties, and on the strength of the on-site e-e repulsion \emph{in primis}.

In conclusion we  have developed a  method to characterize the non-equilibrium steady state of correlated electrons in extended lattices. Combining Floquet and CPT schemes we have defined Floquet spectral functions  that coherently describe how electrons are affected  by their their mutual repulsion and by the interaction with photons. Applying this approach to the 1D Hubbard chain we show that tuning the parameters of the external drive it is possible to turn the Mott insulator into a metal.


%
\end{document}